\documentclass[aps,prl,twocolumn,showpacs,groupedaddress,amssymb,amsfonts,amsmath,apsrev,revsymb]{revtex4-1}

\usepackage{amsmath,amscd}
\usepackage{amssymb,amsfonts}
\usepackage{amsbsy,amsopn,amstext}
\usepackage{graphicx}
\usepackage{revsymb}

\usepackage{dcolumn}
\usepackage{bm}
\usepackage{url}

\usepackage[latin1]{inputenc}

\def\C{{\@QC C}}
\def\@QC#1{\mathpalette{\setbox0=\hbox\bgroup$\rm}%
  {\egroup C$\egroup\rm\rlap{\kern0.4\wd0\vrule
  width 0.05\wd0 height 0.97\ht0 depth -0.01\ht0}%
  #1\bgroup}}

\newcommand{\bra}[1]{\langle#1\rvert}
\newcommand{\ket}[1]{\lvert#1\rangle}
\newcommand{\braket}[2]{\langle#1\mid#2\rangle}
\newcommand{\expval}[3]{\langle#1\rvert#2\lvert#3\rangle}

\newcommand{\bc}{\begin{center}}
\newcommand{\ec}{\end{center}}
\newcommand{\be}{\begin{equation}}
\newcommand{\ee}{\end{equation}}
\newcommand{\bi}{\begin{itemize}}
\newcommand{\ei}{\end{itemize}}

\begin{document}

\title{\textit{Ab-Initio} Molecular Dynamics}
\author{Thomas D. K\"uhne}
\email{kuehne@uni-mainz.de}
\affiliation{Institute of Physical Chemistry and Center for Computational Sciences, Johannes Gutenberg University Mainz, Staudinger Weg 7, D-55128 Mainz, Germany}

\date{\today}

\begin{abstract}
Computer simulation methods, such as Monte Carlo or Molecular Dynamics, are very powerful computational techniques that provide detailed and essentially exact information on classical many-body problems. With the advent of \textit{ab-initio} molecular dynamics, where the forces are computed on-the-fly by accurate electronic structure calculations, the scope of either method has been greatly extended. This new approach, which unifies Newton's and Schr\"odinger's equations, allows for complex simulations without relying on any adjustable parameter. This review is intended to outline the basic principles as well as a survey of the field. Beginning with the derivation of Born-Oppenheimer molecular dynamics, the Car-Parrinello method and the recently devised efficient and accurate Car-Parrinello-like approach to Born-Oppenheimer molecular dynamics, which unifies best of both schemes are discussed. The predictive power of this novel second-generation Car-Parrinello approach is demonstrated by a series of applications ranging from liquid metals, to semiconductors and water. 
This development allows for \textit{ab-initio} molecular dynamics simulations on much larger length and time scales than previously thought feasible. 

\end{abstract}


\maketitle











\section{Introduction}

The geometric increase in performance of computers over the last few decades, together with advances in applied physics and mathematics, has led to the birth of a new way of doing science that is in the intersection of theory as well as experiment. They are referred to as computational sciences and allow for computer experiments under perfectly controllable and reproducible conditions. In this way computer simulations have been very successful in explaining a large variety of physical phenomena and guiding experimental work. In addition it is even possible to predict new phenomena by conducting experiments \textit{in silico} that otherwise would be too difficult, expensive, or simply impossible to perform. However, the by far most rewarding outcome of computer simulations is the invaluable insight they provide into the behavior and the dynamics of a system. The two most common algorithms for such studies are the Monte Carlo (MC) \cite{LandauBinder2009} and Molecular Dynamics (MD) \cite{AlderWainwright1957, Rahman1964} algorithm. The latter is simply the numerical solution of Newton's equation of motion, 
which allows both equilibrium thermodynamic and dynamical properties of a system at finite temperature to be computed. Since it also provides a window' onto the atomic real-time evolution of the atoms, another role of MD is that of a computational microscope. 

One of the most challenging, but very important aspects of MD simulations is the calculation of the interatomic forces. In classical simulations they are computed from empirical potential functions, which have been parametrized to reproduce experimental or accurate \textit{ab-initio} data of small model systems. Even though great strides in elaborating these empirical potentials have been made, often the transferablility to systems or regions of the phase diagram different from the ones to which they have been fitted is restricted. Furthermore, they are not able to simulate with sufficient predictive power chemical bonding processes that take place in many relevant systems. Eventually, some of the most important and interesting phenomena of modern physics and chemistry are intrinsically nonclassical.

Therefore, a first-principles based approach, such as \textit{ab-initio} molecular dynamics (AIMD) \cite{MarxHutter2009}, where the forces are calculated on-the-fly from accurate electronic structure calculations, is very attractive since many of these limitations can in principle be removed. However, the increased accuracy and predictive power of AIMD simulations comes at significant computational cost. For this reason, density functional theory (DFT) \cite{JonesGunnarsson1989} is to date the by far most commonly employed electronic structure theory, but it is important to note that AIMD is a general concept that in principle can be used in conjunction with any electronic structure method. Nevertheless, also the \textit{ab-initio} approach is not without problems - the relevant energy scale is tiny, well below $k_{B}T$, and in particular the attainable length and time scales are still one of its major limitations. 

\section{Molecular Dynamics}

The mathematical task of MD is to evaluate the expectation value $\langle \mathcal{O} \rangle$ of an arbitrary operator $\mathcal{O}(\bm{R}, \bm{P})$ with respect to the Boltzmann distribution
\begin{eqnarray}
  \langle \mathcal{O} \rangle = \frac{\int{d \bm{R} \, d \bm{P} \, \mathcal{O}(\bm{R}, \bm{P}) \, e^{- \beta E(\bm{R}, \bm{P})}}}{\int{d \bm{R} \, d \bm{P} \, e^{- \beta E(\bm{R}, \bm{P})}}}, 
\label{BoltzmannEquation}
\end{eqnarray}
where $\bm{R}$ and $\bm{P}$ are the nuclear positions and momenta, while $\beta = {1}/{k_{B}T}$ is the inverse temperature. The total energy function
\begin{eqnarray} 
  E(\bm{R}, \bm{P}) = \sum_{I=1}^{N}{ \frac{\bm{P}_{I}^{2}}{2 M_{I}} } + \Phi(\bm{R}),
\end{eqnarray}
where the first term denotes the nuclear kinetic energy, $\Phi(\bm{R})$ the potential energy function, $N$ the number of nuclei and $M_{I}$ the corresponding masses. 

One way to to evaluate Eq.~(\ref{BoltzmannEquation}), at least in principle, is to directly solve such a high dimensional integral, whose integrand is very sharply peaked in many dimensions, by an uniform sampling using the MC technique. However, such an algorithm is very inefficient, if it would not be for importance sampling \cite{Metropolis1953}, which satisfies the sufficient detailed balance condition by rejections.

Alternatively, assuming the ergodicity hypothesis, the thermal average $\langle \mathcal{O} \rangle$ can not only be determined as the ensemble average of a MC simulation, but using MD, also as a temporal average
\begin{eqnarray}
  \langle \mathcal{O} \rangle = \lim_{\mathcal{T} \rightarrow \infty} \frac{1}{\mathcal{T}} \int{d t \, \mathcal{O}(\bm{R}(t), \bm{P}(t))}.
  \label{TemporalAverage}
\end{eqnarray}

However, by propagating the classical many-body system in time according to Newton's equation of motion, the nuclei are treated only classical, an usually negligible approximation except for very light atoms or low temperature, where nuclear quantum effects may be important and a quantum formalism such as imaginary-time path integrals is required \cite{ChandlerWolynes1981, ParrinelloRahman1984}.

Similar to MC, within MD some kind of importance sampling is naturally performed by preferentially visiting phase space of low potential energy. Furthermore, as denoted by the additional time dependence in Eq.~(\ref{TemporalAverage}), MD allows for additional insights from the nuclear real-time evolution, at least in an statistical average sense. It is neither the intention, nor possible, to obtain `exact' trajectories by MD due to the infamous Lyapunov instability, which states that slightly perturbed trajectories are intrinsically exponentially diverging with time.

The equipartition theorem
\begin{eqnarray}
  \left\langle \frac{1}{2} \sum_{I=1}^{N} M_{I} \bm{\dot{R}}_{I}^{2} \right\rangle = \frac{3}{2} N k_{B} T, 
  \label{EquipartitionTheorem}
\end{eqnarray}
where the dot indicates time derivative, $k_{B}$ the Boltzmann constant and $T$ the instantaneous temperature, offers an elegant way to bridge the gap between molecular mechanics and thermodynamics. This opens the door to extract a vast variety of relevant static and dynamic, as well as transport properties from a MD simulation.

Nevertheless, any computational resource is finite, which limits the time and length scales accessible by computer simulations. One way to partially bridge the gap between the microscopic size of the simulated system and the macroscopic reality is to introduce periodic boundary conditions (PBC). In this way surface effects are eliminated by effectively simulating an infinite system, albeit with a finite periodicity that is identical to the length $L$ of the simulation cell. As a consequence, only phenomena whose characteristic correlation length is much smaller than $L$ can be simulated, unless a finite-size scaling is applicable \cite{Binder1981}. By similar means only processes whose typical relaxation time is significantly smaller than the simulation time $\mathcal{T}$ can be studied. Even though great strides have been made to extend the accessible time and length scales of AIMD, it is apparent that techniques such as those reviewed here are clearly needed.

\section{An \textit{Ab-Initio} Potential}

In AIMD the interatomic forces $\bm{F}_{I} = - \nabla_{\bm{R}_I} \Phi(\bm{R})$ are determined on-the-fly using first principles electronic structure methods. That means that AIMD is not relying on any adjustable parameter, but only on $\bm{R}$, which constitutes its predictive power. However, finding the antisymmetric ground state eigenfunctions $\ket{\psi_{0}}$ of the corresponding many-body Hamiltonian at each MD step  comes at a significant computational cost, which has to be carefully balanced against the size and sampling requirements of MD.

\subsection{The Many-Body Schr\"odinger Equation}

Applying the so called Born-Oppenheimer approximation \cite{BornOppenheimer1927}, which we have already implicitly assumed in the preceding section, $\Phi(\bm{R})$ can written as
\begin{eqnarray}
  \Phi(\bm{R}) = \expval{\psi_{0}}{\mathcal{H}_{e}(\{ \bm{r}_{i} \}; \bm{R})}{\psi_{0}} + E_{II}(\bm{R}),
  \label{BornOppenheimer}
\end{eqnarray}
where $\mathcal{H}_{e}(\{ \bm{r}_{i} \}; \bm{R})$ is the electronic many-body Hamiltonian, that depend on the electronic coordinates $\{ \bm{r}_{i} \}$ and parametrically on $\bm{R}$. Essentially, the Born-Oppenheimer approximation allows for a \textit{product ansatz} of the total wavefunction consisting of the nuclear and electronic wavefunctions. Due to the large separation of the nuclear and electronic masses, the electrons can be expected to be in instantaneous equilibrium with the much heavier nuclei, so that the electronic subsystem can be treated independently at constant $\bm{R}$, hence the parametric dependence of $\mathcal{H}_{e}(\{ \bm{r}_{i} \}; \bm{R})$. Nevertheless, we are left with the formidable task to solve the electronic, nonrelativistic, time-independent, many-body Schr\"odinger equation 
\begin{eqnarray}
  \mathcal{H}_{e}(\{ \bm{r}_{i} \}; \bm{R}) \psi_{0}(\{ \bm{r}_{i} \}) = \varepsilon_{0}(\bm{R}) \psi_{0}(\{ \bm{r}_{i} \})
  \label{SchroedingerEq}
\end{eqnarray}
that is a high-dimensional eigenvalue problem, with eigenfunctions $\psi_{0}(\{ \bm{r}_{i} \})$ and eigenvalues $\varepsilon_{0}(\bm{R})$, respectively.

To visualize the complexity of Eq.~(\ref{SchroedingerEq}) let us consider the following \textit{Gedankenmodell} to represent the solution $\psi_{0}(\{ \bm{r}_{i} \})$ on a real-space grid, where each coordinate is discretized by as less as 100 mesh points. Ignoring spin and taking $\psi_{0}(\{ \bm{r}_{i} \})$ to be real, for $N_{e}$ electrons $10^{6N_{e}}$ grid points are required, so that the solution of a single Si atom would require more grid points than the number of electrons in the whole universe, not to mention solving such a large non-linear eigenvalue problem. 

\subsection{Density Functional Theory}

Fortunately, this curse of dimensionality can be ingeniously bypassed by the use of DFT, which is based on two celebrated papers of Hohenberg, Kohn and Sham \cite{HohenbergKohn1964, KohnSham1965}. The former, the so called Hohenberg-Kohn (HK) theorem, proves the existence of an one-to-one mapping between the ground state density $\rho_{0}(\bm{r})$ and an external potential $v(\bm{r})$. Since the electronic density $\rho(\bm{r}) = \int{\bm{r}_{2} \cdot \cdot \cdot \int{\bm{r}_{N_{e}} | \psi(\{ \bm{r}_{i} \}) |^{2} }}$ depends on just 3 electronic degrees of freedom, it is in DFT designated as the principal quantity instead of the more complex $3N_{e}$-dimensional many-body wavefunction. Hence, the nondegenerate ground state wavefunction $\psi_{0}(\{ \bm{r}_{i} \}) = \psi[\rho_{0}(\bm{r})]$ and likewise $\mathcal{H}_{e}[\rho_{0}(\bm{r})]$ are both unique functionals of $\rho_{0}(\bm{r})$, just as the ground state energy $E_{0} = E^{\text{DFT}}[\rho_{0}(\bm{r})] = \expval{\psi[\rho_{0}(\bm{r})]}{\mathcal{H}_{e}[\rho_{0}(\bm{r})]}{\psi[\rho_{0}(\bm{r})]}$. 
The latter obeys the variational property 
\begin{eqnarray} 
  E^{\text{DFT}}[\rho_{0}] = \expval{\psi_{0}}{\mathcal{H}_{e}}{\psi_{0}} \leq \expval{\psi^{\prime}}{\mathcal{H}_{e}}{\psi^{\prime}} = E^{\text{DFT}}[\rho^{\prime}],
  \label{VariationalProperty}
\end{eqnarray}
for which equality holds if and only if $\rho_{0} = \rho^{\prime}$.
Thus, Eq.~(\ref{SchroedingerEq}) can not only be solved by iteratively diagonalizing $\mathcal{H}_{e}[\rho]$ within a self-consistent field (SCF) procedure, but also by minimizing the quantum expectation value of $\mathcal{H}_{e}[\rho]$, i.e. 
%
\begin{eqnarray}
  E^{DFT}[\rho_{0}] &=& \min_{\psi}{ \expval{\psi}{\mathcal{H}_{e}}{\psi} } = \min_{\rho}{ \expval{\psi[\rho]}{\mathcal{H}_{e}[\rho]}{\psi[\rho]} } \nonumber \\
  &=& \min_{\rho}{ E^{\text{DFT}}[\rho] }.
  \label{ConstraintMinimization}
\end{eqnarray}

In principle, the minimization has to be performed under the constraint that $\rho(\bm{r})$ is $N$-representable, i.e. that it is arising from an antisymmetric N-body wavefunction $\psi(\{ \bm{r}_{i} \})$. Luckily, this had been solved, and it can be demonstrated that any single-particle density can be written in terms of an antisymmetric many-body wavefunction \cite{Gilbert1975, Harriman1981}. On the contrary, for the $v$-representability problem, which states that $\rho(\bm{r})$ is the ground state density of a local potential $v(\bm{r})$, no such general solution is known. The HK theorem just guarantees that there cannot be more than one potential for each density, but does not exclude the possibility that there is no potential realizing that density. It is only known for discretized systems that every density is interacting ensemble $v$-representable. Interestingly, the constructive proof of Levy and Lieb \cite{Levy1982, Lieb1983} shows that for an arbitrary interacting system $v$-representability is not required for the proof of the HK theorem.

For the sake of simplicity, in the following I will throughout assume atomic units and confine myself to the physical relevant coulomb system, for which
\begin{subequations}
\begin{eqnarray}
  \mathcal{H}_{e} &=& \frac{1}{2} \sum_{i=1}^{N_{e}}{\nabla_{i}^{2}} + \sum_{i<j}^{N_{e}}{\frac{1}{| \bm{r}_{i} - \bm{r}_{j} |}} + \sum_{I, i}^{N, N_{e}}{ \frac{Z_{I}}{| \bm{R}_{I} - \bm{r}_{i} |} } \\ &=& \hat{T} + \hat{U} + \hat{V}, 
  \label{Helec}
\end{eqnarray}
\end{subequations}
where $Z_{I}$ is the proton number, $\hat{T}$ the kinetic energy operator of the electrons, while $\hat{U}$ is the electron-electron interaction and $\hat{V} = \sum_{i}v(\bm{r}_{i})$ the electron-ion operator. The former two operators are universal and independent of the system, while the latter is system dependent, or nonuniversal. DFT explicitly recognizes that it is indeed the potential $v(\bm{r})$, which distinguishes nonrelativistic Coulomb systems and offers a prescription how to deal with $\hat{T}$ and $\hat{U}$ once and for all. 
Therefore, even at this stage, using nothing but the HK theorem, DFT is already of some practical use without having to solve the many-body Schr\"odinger equation and without having to make a single-particle approximation by writing 
\begin{eqnarray}
  E^{\text{DFT}}[\rho(\bm{r})] = T[\rho(\bm{r})] + U[\rho(\bm{r})] + V[\rho(\bm{r})], 
  \label{ETF}
\end{eqnarray}
where $T[\rho(\bm{r})]$ is the kinetic, $U[\rho(\bm{r})]$ the electron-electron and $V[\rho(\bm{r})]$ the electron-ion interaction energy as a functional of $\rho(\bm{r})$. 
In principle it should be even possible to calculate all observables, since the HK theorem guarantees that they are all functionals of $\rho(\bm{r})$. However, although the functional $T[\rho(\bm{r})] + U[\rho(\bm{r})]$ is universal for all systems of $N_{e}$ electrons, its explicit form is unfortunately unknown, so that physical sound and reliable approximations are indispensable. 

In the so called Thomas-Fermi (TF) approximation \cite{Thomas1927, Fermi1927}, the full electron-electron interaction energy is approximated by the Hartree energy
\begin{eqnarray}
  U[\rho(\bm{r})] \approx U_{\text{H}}[\rho(\bm{r})] = \frac{1}{2} \int{ d\bm{r} \, \int{ d\bm{r}^{\prime} \, \frac{\rho(\bm{r}) \rho(\bm{r}^{\prime})}{| \bm{r} - \bm{r}^{\prime} |} } }
  \label{Ehartree}
\end{eqnarray}
that is the classical electrostatic interaction energy between the electron densities $\rho(\bm{r})$ and $\rho(\bm{r}')$. In addition, the kinetic energy is approximated as 
\begin{eqnarray}
  T[\rho(\bm{r})] \approx \int{ d\bm{r} \, t^{hom}(\rho(\bm{r})) } = T^{\text{LDA}}[\rho(\bm{r})], 
  \label{DFTkinE}
\end{eqnarray}
where $t^{hom}(\rho(\bm{r}))$ is the kinetic-energy density of a homogeneous interacting system, which is also known as the local-density approximation (LDA) to $T[\rho(\bm{r})]$. Since that the explicit form of $t^{hom}(\rho(\bm{r}))$ is only known for a homogeneous non-interacting system, $t^{hom}(\rho(\bm{r}))$ is further estimated by the single-particle approximation $t_{s}^{hom}(\rho(\bm{r}))$, i.e. 
\begin{eqnarray}
  T^{\text{LDA}}[\rho(\bm{r})] \approx \int{ d\bm{r} \, t_{s}^{hom}(\rho(\bm{r})) } = T_{s}^{\text{LDA}}[\rho(\bm{r})], 
  \label{TFkinE}
\end{eqnarray}
where 
\begin{eqnarray}
  t_{s}^{hom}(\rho(\bm{r})) = \frac{3}{10} (3 \pi^{2})^{2/3} \rho(\bm{r})^{5/3}.
\end{eqnarray}

In the end the TF energy functional 
\begin{eqnarray}
  E^{\text{TF}}[\rho(\bm{r})] &=& T_{s}^{\text{LDA}}[\rho(\bm{r})] + U_{\text{H}}[\rho(\bm{r})]  + V[\rho(\bm{r})] \nonumber \\
  &\approx& E^{\text{DFT}}[\rho(\bm{r})]
  \label{TFenergy}
\end{eqnarray}
implies not only the single-particle approximation to the full electron-electron interaction, but also the single-particle mean-field approximation $T_{s}^{\text{LDA}}[\rho(\bm{r})]$ to the exact kinetic energy of the inhomogeneous interacting system. As a consequence, all many-body correlation effects are at this point neglected. 

However, the HK theorem predicates that these many-body correlation effects are again a functional of $\rho(\bm{r})$ and is known as the exchange and correlation (XC) functional. The addition of an approximation to the exact XC functional results in a formally exact theory, which is referred to as orbital-free DFT \cite{SmargiassiMadden1994}. It is therefore important to recognize, that the HK theorem is nothing but the formal exactification of the TF approximation. 

Similarly, the Kohn-Sham (KS) \cite{KohnSham1965} scheme can be considered as the exactification of the self-consistent Hartree equations (HE) \cite{Hartree1928} that differs from the TF approximation only in the kinetic energy, which is typically rather larger. 
Fortunately, it is not necessary to approximate the kinetic energy of the inhomogeneous non-interacting system, as is done in TF and orbital-free DFT, but is known exactly, even though only in terms of an explicit orbital functional, i.e. as an implicit density functional
%
\begin{subequations}
\begin{eqnarray}
  T_{s}[\rho(\bm{r})] &=& - \frac{1}{2} \sum_{i=1}^{N_{e}} \int{ d\bm{r} \, \psi_{i}^{*}(\bm{r}) \nabla^{2} \psi_{i}(\bm{r}) } \nonumber \\
  &=& T_{s} \left[ {\{ \psi_{i}[\rho(\bm{r})] \}} \right].
  \label{FTkinE}
\end{eqnarray}
\end{subequations}
Here the fictitious single-particle wavefunctions, or simply KS orbitals, are denoted as $\psi_{i}(\bm{r})$. As we will see immediately they are eigenfunctions of a fictitious system, known as the KS system. It therefore should be noted that they differ from the single-particle orbitals used in wavefunction based methods and have no strict physical meaning, with two notable exceptions: at the presence of the exact XC functional for the special case of an isolated system with $v(\infty) = 0$ (i) the highest occupied eigenvalue $\varepsilon_{N}$ can be shown to be the negative of the exact, many-body, first ionization potential including relaxation effects, and (ii) that the lowest unoccupied eigenvalue $\varepsilon_{N+1}$ is the negative of the electron affinity. Beside these two exceptions, only the density has a real physical meaning and 
can be written in terms of $\psi_{i}(\bm{r})$ as 
\begin{eqnarray}
  \rho(\bm{r}) = \sum_{i=1}^{N_{occ}}{ f_{i} \psi_{i}(\bm{r}) \psi_{i}^{*}(\bm{r}) },
  \label{ChargeDensity}
\end{eqnarray}
where $N_{occ}$ is the number of occupied orbitals and $f_{i}$ the occupation number of state $i$, so that 
\begin{eqnarray}
  \sum_{i=1}^{N_{occ}}{ f_{i} } = N_{e}.
  \label{Nelec}
\end{eqnarray}
Therewith, the KS energy functional is simply given by
\begin{subequations}
\begin{eqnarray}
  E^{\text{KS}}[\rho(\bm{r})] &=& E^{\text{KS}} \left[ {\{ \psi_{i}[\rho(\bm{r})] \}} \right] = T_{s} \left[ {\{ \psi_{i}[\rho(\bm{r})] \}} \right] \nonumber \\ 
  &+& U_{\text{H}}[\rho(\bm{r})] + V[\rho(\bm{r})] + E_{\text{XC}}[\rho(\bm{r})] \label{KSenergyA} \\
  &=&-\frac{1}{2} \sum_{i=1}^{N} f_{i} \int{ d\bm{r} \, \psi_{i}^{*}(\bm{r}) \nabla^{2} \psi_{i}(\bm{r}) } \nonumber \\
  &+& \frac{1}{2} \int{ d\bm{r} \int d\bm{r}^{\prime} \, \frac{ \rho(\bm{r}) \rho(\bm{r}^{\prime}) }{ \lvert \bm{r} - \bm{r}^{\prime} \rvert } } \nonumber \\
  &+& \int{ d\bm{r} \, v_{\text{ext}}(\bm{r}) \rho(\bm{r}) } + E_{\text{XC}}[\rho(\bm{r})], \qquad \label{KSenergyB}
\end{eqnarray}
\end{subequations}
where $E_{\text{XC}}[\rho(\bm{r})] = (T[\rho(\bm{r})] - T_{s} \left[ {\{ \psi_{i}[\rho(\bm{r})] \}} \right]) + (U[\rho(\bm{r})] - U_{\text{H}}[\rho(\bm{r})])$ is the already mentioned and apparently unknown XC energy functional and $v_{\text{ext}}(\bm{r}) = \delta V[\rho(\bm{r})] / \delta \rho(\bm{r})$ the so called external potential, which is not self-consistently generated from the electron-electron interaction, but external to the electronic system. This definition also shows that a significant of part $E_{\text{XC}}[\rho(\bm{r})]$ is due to correlation effects of the kinetic energy, that is known explicitly only in terms of the reduced 2-particle density matrix \cite{DreizlerGross1990}.

Due to the fact that $T_{s}\left[ {\{ \psi_{i}[\rho(\bm{r})] \}} \right]$ is now an explicit orbital functional it is not straightforward to directly minimize Eq.~(\ref{KSenergyA}). Instead, it is much more appropriate to make $E^{\text{KS}} \left[ {\{ \psi_{i}[\rho(\bm{r})] \}} \right]$ stationary by the following Euler-Lagrange equation: 
\begin{subequations}
\begin{eqnarray}
  0 &=& \frac{\delta E^{\text{KS}}[\rho(\bm{r})]}{\delta \rho(\bm{r})} = \frac{\delta E^{\text{KS}} \left[ {\{ \psi_{i}[\rho(\bm{r})] \}} \right]}{\delta \rho(\bm{r})} \\
  &=& \frac{T_{s} \left[ {\{ \psi_{i}[\rho] \}} \right]}{\delta \rho(\bm{r})} + \frac{\delta U_{\text{H}}[\rho]}{\delta \rho(\bm{r})} + \frac{\delta V[\rho]}{\delta \rho(\bm{r})} + \frac{\delta E_{\text{XC}}[\rho]}{\delta \rho(\bm{r})} \qquad \\ 
  &=& \frac{T_{s} \left[ {\{ \psi_{i}[\rho] \}} \right]}{\delta \rho(\bm{r})} + v_{\text{H}}(\bm{r}) + v_{ext}(\bm{r}) + v_{\text{XC}}(\bm{r}), \label{KS-EulerLagrange}
\end{eqnarray}
\end{subequations}
where $v_{\text{H}}(\bm{r}) = 1/2 \int{d\bm{r}' \rho(\bm{r}') / |\bm{r} - \bm{r}'|}$ is the Hartree potential and $v_{\text{XC}}(\bm{r}) = \delta E_{\text{XC}}[\rho(\bm{r})] / \delta \rho(\bm{r})$ the XC potential. The corresponding Euler-Lagrange equation of the noninteracting system within the potential $v_{s}(\bm{r})$ reads as
\begin{eqnarray}
  0= \frac{T_{s} \left[ {\{ \psi_{i}[\rho] \}} \right]}{\delta \rho(\bm{r})} + v_{s}(\bm{r}), \label{SingleParticleEulerLagrange}
\end{eqnarray}
where $v_{\text{H}}(\bm{r})$ and $v_{\text{XC}}(\bm{r})$ are vanishing in the absence of interactions. Comparing Eq.~(\ref{KS-EulerLagrange}) with (\ref{SingleParticleEulerLagrange}) immediately suggest the possibility to solve the former by a fictitious single-particle equation
%
\begin{eqnarray}
  \left( - \frac{1}{2} \nabla^{2} + v_{s}^{\text{KS}}(\bm{r}) \right) \psi_{i}(\bm{r}) &=& \varepsilon_{i} \psi_{i}(\bm{r}), \label{SingleParticleSE}
\end{eqnarray}
%
where $v_{s}^{\text{KS}}(\bm{r}) = v_{\text{H}}(\bm{r}) + v_{\text{ext}}(\bm{r}) + v_{\text{XC}}(\bm{r})$ is the effective potential of an artificial system, such that the ground state density and therewith the energy equals those of the true interacting many-body system. This particular system is therefore called KS system, its effective potential KS potential and the resulting set of self-consistent equations are referred to as KS equations:
%
\begin{eqnarray}
  \left( - \frac{1}{2} \nabla^{2} 
  + v_{\text{H}}(\bm{r}) + v_{\text{ext}}(\bm{r}) + v_{\text{XC}}(\bm{r}) \right) \psi_{i}(\bm{r}) &=& \varepsilon_{i} \psi_{i}(\bm{r}) \qquad \nonumber \\
  \sum_{i=1}^{N_{occ}}{ f_{i} \psi_{i}(\bm{r}) \psi_{i}^{*}(\bm{r}) } = \rho_{s}(\bm{r}) &\equiv& \rho(\bm{r}) \label{KSeq} 
\end{eqnarray}

In other words, to enable genuine calculations the KS scheme systematically maps the full interacting many-body problem, with $\hat{U}$, onto an equivalent fictitious single-body problem, with an effective potential operator $\hat{V}_{\text{KS}} = \hat{U}_{s} + \hat{V}_{H} + \hat{V}_{XC}$, but without $\hat{U}$ \cite{Capelle2006}:
%
\begin{equation*}
\begin{CD}
  \text{TF} @>{E_{\text{XC}}[\rho(\bm{r})]}>> \text{HK} \\
  @V{T_{s} \left[ {\{ \psi_{i}[\rho(\bm{r})] \}} \right]}VV @VV{T_{s} \left[ {\{ \psi_{i}[\rho(\bm{r})] \}} \right]}V \\
  \text{HE} @>>{E_{\text{XC}}[\rho(\bm{r})]}> \text{KS}
\end{CD}
\end{equation*}

At self-consistency it is possible to express $E_{0}^{\text{KS}}[\rho(\bm{r})]$ in terms of the single-particle KS eigenvalues $\varepsilon_{i}$. Due to the fact that they are not the eigenvalues of the interacting many-body system, but of the fictitious non-interacting KS system instead, $E_{0}^{\text{KS}}[\rho(\bm{r})]$ is merely the sum of $\varepsilon_{i}$, but rather 
\begin{eqnarray}
  E_{0}^{\text{KS}}[\rho(\bm{r})] &=& \sum_{i}^{N_{occ}} {f_{i} \varepsilon_{i}} - \frac{1}{2} \int{ d\bm{r} \int d\bm{r}^{\prime} \, \frac{ \rho(\bm{r}) \rho(\bm{r}^{\prime}) }{ \lvert \bm{r} - \bm{r}^{\prime} \rvert } } \nonumber \\
  &-& \int{ d\bm{r} \, v_{\text{XC}}(\bm{r}) \rho(\bm{r}) } + E_{\text{XC}}[\rho(\bm{r})].
  \label{SumEval}
\end{eqnarray}
To so called double counting terms on the right hand side of Eq.~(\ref{SumEval}) are a manifestation that the whole is more than the sum of its individual parts. 

\subsection{The Exchange and Correlation Functional}

In the previous subsection DFT has been outlined as a formally exact theory, presuming that the exact XC functional is known. Unfortunately, except for the uniform electron gas \cite{CeperleyAlder1980}, this is not the case and one has to resort to more or less accurate approximations. 
On this account the following break-up is particularly convenient:
\begin{eqnarray}
  E_{\text{XC}}[\rho(\bm{r})] &=& \frac{1}{2} \int{ d\bm{r} \, \int{ d\bm{r}^{\prime} \, \frac{\rho(\bm{r}) \rho_{\text{XC}}(\bm{r}, \bm{r}^{\prime}) }{| \bm{r} - \bm{r}^{\prime} |} } } \nonumber \\
  &=& E_{\text{X}}[\rho(\bm{r})] + E_{\text{C}}[\rho(\bm{r})],
  \label{XCbreak}
\end{eqnarray}
where $E_{\text{X}}[\rho(\bm{r})]$ is the exchange energy due to the Pauli repulsion, $E_{\text{C}}[\rho(\bm{r})]$ the electron correlation energy and $\rho_{\text{XC}}(\bm{r}, \bm{r}^{\prime}) = \rho_{\text{X}}(\bm{r}, \bm{r}^{\prime}) + \rho_{\text{C}}(\bm{r}, \bm{r}^{\prime})$ the XC hole. The former therefore corresponds to the energy lowering due to the antisymmetry requirement on the wavefunction of a fermionic system. Moreover, similar to $T_{s}\left[ {\{ \psi_{i}[\rho(\bm{r})] \}} \right]$, it is also possible to exactly calculate $E_{\text{X}}[\rho(\bm{r})]$ in terms of an explicit orbital and implicit density functional
\begin{eqnarray}
   E_{\text{X}} [\rho(\bm{r})] &=& - \frac{1}{2} \sum_{i, j}{\int{ d\bm{r} \, \int{ d\bm{r}^{\prime} \, \frac{\psi_{i}^{*}(\bm{r}) \psi_{j}^{*}(\bm{r})^{\prime} \psi_{i}(\bm{r}) \psi_{j}(\bm{r})^{\prime}}{| \bm{r} - \bm{r}^{\prime} |} } }} \nonumber \\ 
  &=& E_{\text{X}} \left[ {\{ \psi_{i}[\rho(\bm{r})] \}} \right]. 
  \label{EXX}
\end{eqnarray}
This is known as the Hartree-Fock exchange energy, though here with KS orbitals. Unfortunately, the nonlocal nature of Eq.~(\ref{EXX}) comes at a considerable computational burden to solve four-center integrals, which in particular for periodic systems is substantially more costly than the commonly employed local or semi-local approximations to the exact XC functional. However, the development of ingenious screening and auxiliary density matrix techniques is, in the near future, expected to facilitate the routine inclusion of exact exchange in AIMD simulations of large periodic systems \cite{TodorovaHutterMundy2006, Guidon2008, Guidon2009, Guidon2010}. 
Apart from that, the correlation energy accounts for the additional energy lowering, since electrons with opposite spins also avoid each other. Yet, contrary to the exchange part, no exact expression for $E_{\text{C}}[\rho(\bm{r})]$ is known, neither in terms of orbitals nor densities. 

Obviously, DFT would be of little use if one had to know $E_{\text{XC}}[\rho(\bm{r})]$ exactly, but luckily it is usually energetically substantially smaller than each the remaining terms that are known. One can thus hope that reasonable simple approximations to $E_{\text{XC}}[\rho(\bm{r})]$, will still allow for qualitatively correct estimates of $E_{0}[\rho(\bm{r})]$, without relying on additional adjustable parameters.

\section{\textit{Ab-Initio} Molecular Dynamics}

In the following, we will assume that the potential energy function is calculated on-the-fly using DFT, so that $\Phi(\bm{R}) = E[\{ \psi_{i} \}; \bm{R}] = E^{\text{KS}} \left[ \{ \psi_{i}[\rho(\bm{r})] \}; \bm{R} \right] + E_{II}(\bm{R})$. In any case, AIMD \cite{CarParrinello1985, Payne1992, Tuckerman2002, MarxHutter2009} comes in two fundamental flavors, which are outlined in this section.

\subsection{Born-Oppenheimer Molecular Dynamics}

In Born-Oppenheimer MD (BOMD) the potential energy $E \left[ \{ \psi_{i} \}; \bm{R} \right]$ is minimized at every MD step with respect to $\{ \psi_{i}({\bm{r}}) \}$ under the holonomic orthonormality constraint $\braket{\psi_{i}(\bm{r})}{\psi_{j}(\bm{r})} = \delta_{ij}$. 
This leads to the following Lagrangian:
\begin{eqnarray}
  \mathcal{L}_{\text{BO}} \left( \{\psi_{i}\}; \bm{R}, \dot{\bm{R}} \right) &=& \frac{1}{2} \sum_{I=1}^{N}{M_{I}\dot{\bm{R}}_{I}^{2}} - \min_{ \{\psi_{i}\} } E \bigl[ \{ \psi_{i} \}; \bm{R} \bigr] \nonumber \\
  &+& \sum_{i,j} { \Lambda_{ij} \left( \braket{\psi_{i}}{\psi_{j}} - \delta_{ij} \right) }, \label{BO_Lagrangian}
\end{eqnarray}
where $\bm{\Lambda}$ is a Hermitian Lagrangian multiplier matrix. By solving the corresponding Euler-Lagrange equations
\begin{subequations}
\begin{eqnarray}
  \frac{d}{dt} \frac{\partial \mathcal{L}}{\partial \dot{\bm{R}}_{I}} &=& \frac{\partial \mathcal{L}}{\partial \bm{R}_{I}} \label{ElectronEulerLagrangian} \\
  \frac{d}{dt} \frac{\partial \mathcal{L}}{\partial \bra{\dot{\psi}_{i}}} &=& \frac{\partial \mathcal{L}}{\partial \bra{\psi_{i}}} \label{IonEulerLagrangian}
\end{eqnarray}
\end{subequations}
one obtains the associated equations of motion (EOM)
\begin{subequations}
\begin{eqnarray}
  M_{I} \Ddot{\bm{R}}_{I} &=& -\nabla_{\bm{R}_{I}} \left[ \min_{ \{\psi_{i}\} } E \bigl[ \{ \psi_{i} \}; \bm{R} \bigr] \biggm|_{ \{ \braket{\psi_{i}}{\psi_{j}} = \delta_{ij} \} } \right] \nonumber \\
  &=& - \frac{\partial E}{\partial \bm{R}_{I}} + \sum_{i,j} \Lambda_{ij} \frac{\partial}{\partial \bm{R}_{I}} \braket{\psi_{i}}{\psi_{j}} \nonumber \\ 
  &-& 2 \sum_{i} \frac{\partial \bra{\psi_{i}}}{\partial \bm{R}_{I}} \left[ \frac{\delta E}{\delta \bra{\psi_{i}}} - \sum_{j} \Lambda_{ij} \ket{\psi_{j}} \right] \label{IonBOMD_EOM} \\
  0 &\lesssim& - \frac{\delta E}{\delta \bra{\psi_{i}}} + \sum_{j} \Lambda_{ij} \ket{\psi_{j}} \nonumber \\
  &=& -\hat{H}_{e} \bra{\psi_{i}} + \sum_{j} \Lambda_{ij} \ket{\psi_{j}} \label{ElectronBOMD_EOM}
\end{eqnarray}
\end{subequations}

The first term on the righ-hand side (RHS) of Eq.~(\ref{IonBOMD_EOM}) is the so called Hellmann-Feynman force \cite{Hellmann1937, Feynman1939}. The second term that is denoted as Pulay \cite{Pulay1969}, or wavefunction force $F_{\text{WF}}$, is a constraint force due to the holonomic orthonormality constraint, and is nonvanishing if and only if the basis functions $\phi_{j}$ explicitly depend on $\bm{R}$. The final term stems from the fact that, independent of the particular basis set used, there is always an implicit dependence on the atomic positions through the expansion coefficient $c_{ij}(\bm{R})$ within the common linear combination of atomic orbitals $\phi_{j}$:
\begin{eqnarray}
  \psi_{i}(\bm{R}) = \sum_{j}{c_{ij}(\bm{R}) \phi_{j}}
  \label{LCAO}
\end{eqnarray}
The factor 2 in Eq.~(\ref{IonBOMD_EOM}) stems from the assumption that the KS orbitals are real, an inessential simplification. Nevertheless, the whole term vanishes whenever $\psi_{i}(\bm{R})$ is an eigenfunction of the Hamiltonian within the subspace spanned by the not necessarily complete basis set \cite{AlmloefHelgaker1981, Scheffler1985}. Note, that this is a much weaker condition than the original Hellmann-Feynman theorem, which we hence haven't availed throughout the derivation, except as an eponym for the first RHS term of Eq.~(\ref{IonBOMD_EOM}). However, as the KS functional is nonlinear, eigenfunctions of its Hamiltonian $\hat{H}_{e}$ are only obtained at exact self-consistency, which is why the last term of Eq.~(\ref{IonBOMD_EOM}) is also referred to as non-self-consistent force $F_{\text{NSC}}$. Unfortunately, in any numerical calculation this can not be assumed and results in immanent inconsistent forces as well as the inequality of Eq.~(\ref{ElectronBOMD_EOM}). Neglecting either $F_{\text{WF}}$ or $F_{\text{NSC}}$, i.e. applying the Hellmann-Feynman theorem to a non-eigenfunction leads merely to a perturbative estimate of the generalized forces \cite{BendtZunger1983}
\begin{eqnarray}
  F = F_{\text{HF}} + F_{\text{WF}} + F_{\text{NSC}},
  \label{GeneralizedForce}
\end{eqnarray}
which, contrary to the energies, depends just linearly on the error in the electronic charge density. That is why it is much more exacting to calculate accurate forces than total energies.

Nevertheless, as a corollary of the BO approximation, the electronic and nuclear subsystems are fully decoupled from each other. Due to the adiabatic separation there are no additional restrictions on the maximum permissible integration time step, which hence can be chosen up to the nuclear resonance limit. In fact, this actually holds irrespective of the band gap, so as to, at least in principle, even metals can be straightforwardly treated.

\subsection{Car-Parrinello Molecular Dynamics}

In Car-Parrinello MD (CPMD) \cite{CarParrinello1985} a coupled electron-ion dynamics is performed, wherein the electronic degrees of freedom are added to the Lagrangian as classical ones: 
\begin{eqnarray}
  \mathcal{L}_{\text{CP}} \bigl( \{\psi_{i}\}; \bm{R}, \dot{\bm{R}} \bigr) &=& \frac{1}{2} \mu \sum_{i} \braket{\dot{\psi}_{i}}{\dot{\psi}_{i}} + \frac{1}{2} \sum_{I=1}^{N}{M_{I}\dot{\bm{R}}_{I}^{2}} \nonumber \\
  &-& E \bigl[ \{ \psi_{i} \}; \bm{R} \bigr] \nonumber \\
  &+& \sum_{i,j} {\Lambda_{ij} \bigl( \braket{\psi_{i}}{\psi_{j}} - \delta_{ij} \bigr)}, \label{CPlagrangian}
\end{eqnarray}
where the electronic degrees of freedom inhere an artificial inertia commonly known as the fictitious mass parameter $\mu$, whose units is energy $\times$ time$^{2}$, as well as orbital velocities $\{\dot{\psi}_{i}\}$. 
Once again, applying the Euler-Lagrange equations of Eq.~(\ref{ElectronEulerLagrangian}) and (\ref{IonEulerLagrangian}) leads to the following EOM:
\begin{subequations}
\begin{eqnarray}
  M_{I} \Ddot{\bm{R}}_{I} &=& -\nabla_{\bm{R}_{I}} \left[ E \bigl[ \{ \psi_{i} \}; \bm{R} \bigr] \biggm|_{ \{ \braket{\psi_{i}}{\psi_{j}} = \delta_{ij} \} } \right] \nonumber \\
  &=& - \frac{\partial E}{\partial \bm{R}_{I}} + \sum_{i,j} \Lambda_{ij} \frac{\partial}{\partial \bm{R}_{I}} \braket{\psi_{i}}{\psi_{j}} \label{IonCPMD_EOM} \\
  \mu \Ddot{\psi}_{i}(\bm{r}, t) &=& - \frac{\delta E}{\delta \bra{\psi_{i}}} + \sum_{j} \Lambda_{ij} \ket{\psi_{j}} \nonumber \\
  &=& - \hat{H}_{e} \bra{\psi_{i}} + \sum_{j} \Lambda_{ij} \ket{\psi_{j}}, \label{ElectronCPMD_EOM}
\end{eqnarray}
\end{subequations}
where $-\delta E / \delta \bra{\psi_{i}}$ are the electronic forces to propagate the electronic degrees of freedom in time within a fictitious Newtonian dynamics along the nuclei. At variance to BOMD, the computational cost to compute the nuclear forces in each AIMD step is now much reduced since no SCF cycle is required to ensure that they are consistent with the instantaneous energy, and to force the electrons to adiabatically follow the nuclei. By contrast, in BOMD this can only be accomplished via the full constraint minimization $\min_{ \{\psi_{i}\} } E \left[ \{ \psi_{i} \}; \bm{R} \right] \bigm|_{ \{ \braket{\psi_{i}}{\psi_{j}} = \delta_{ij} \} }$ within Eq.~(\ref{IonBOMD_EOM}), i.e. by constraining the orbitals to their instantaneous ground state. 

However, in order to ensure the aforementioned adiabatic energy-scale separation of the nuclear and the electronic degrees of freedom and to prevent energy transfer between them, the highest ionic phonon frequency $\omega_{I}$ has to be much smaller than its lowest electronic analog $\omega_{e}$. The latter has been found to behave like  
\begin{eqnarray}
  \omega_{e} \propto \sqrt{\frac{\Delta E_{gap}}{\mu}}, \label{PastoreBuda}
\end{eqnarray}
where $\Delta E_{gap}$ is the KS single-particle energy gap \cite{PastoreBuda1991}.
Hence, the maximum joint integration time step ${\Delta t}_{max} < 1 / \omega_{e}$ depends on the inertia like $\sqrt{\mu}$. The same also holds for the deviation from the BO surface 
\begin{eqnarray}
  | \psi_{\mu}(\bm{r}, t) - \psi_{0}(\bm{r}, t) | \leq C \sqrt{\mu},
  \label{BornemannSchuette}
\end{eqnarray}
where $\psi_{\mu}(\bm{r}, t)$ is the CP wavefunction as propagated by Eq.~(\ref{ElectronCPMD_EOM}), which is energetically always slightly above the electronic ground state, whose wavefunction is denoted by $\psi_{0}(\bm{r}, t)$, whereas $C$ an inessential constant \cite{BornemannSchuette1998}. In other words, although physically meaningless, $\mu$ acts as a continuous slider to trade computational efficiency for accuracy in terms of deviation from the instantaneous BO surface. Nevertheless, if the latter corresponds merely to an approximate constant shift, the nuclear forces are thus not affected. 

As already said, $\mu$ has to be chosen small enough in order that $\omega_{I} << \omega_{e}$, from which it follows that the high-frequency oscillations of Eq.~(\ref{ElectronCPMD_EOM}) vanishes on ionic time scales, i.e. $\langle \Ddot{\psi}_{i} \rangle \simeq 0$. Therefore, similar to Ehrenfest dynamics \cite{Ehrenfest1927}, the total derivative of the instantaneous, rather than the fully minimized, expectation value $\expval{\Psi_{0}}{\hat{H}_{e}}{\Psi_{0}}$ of the electronic Hamiltonian yields the correct consistent forces. This implies that owing to the absence of necessity to fully minimize $E \left[ \{ \psi_{i} \}; \bm{R} \right] \bigm|_{ \{ \braket{\psi_{i}}{\psi_{j}} = \delta_{ij} \} }$, but rather to simply evaluate it for any give time step, $F_{NSC}$ is identical to zero by its very definition. Provided that $\mu$ is small enough, the constant of motion is strictly conserved and the errors in the forces negligible, in particular if the ionic masses are renormalized by a constant mass tensor \cite{Bloechl1992, TangneyScandolo2002}. As such, CPMD combines most of the advantages of BOMD and Ehrenfest MD. In the latter, the electronic degrees of freedom are evolved according to the time-dependent Schr\"odinger equation that, in the limit $\Delta t \rightarrow 0$, corresponds to a unitary propagation. Thus, unlike BOMD, there is no need to repeatedly diagonalize $\mathcal{H}_{e}$ or, equivalently, iteratively minimize $E \left[ \{ \psi_{i} \}; \bm{R} \right]$, just as in CPMD. Furthermore, the unitary transformation within Ehrenfest MD entails that wavefunctions will remain orthonormal to each other, even though at the price that the joint integration timestep must be is tiny and is commonly confined to a few attoseconds. On the contrary, in CPMD the joint integration time step is typically allowed to be roughly two orders of magnitude larger than that, which is only around one order of magnitude short of BOMD. However, due to the finite accuracy of any integrator, in CPMD the holonomic orthonormality constraint of the orbitals has to be explicitly enforced. Due to the fact that this is more than compensated by the larger integration time step of CPMD, Ehrenfest MD is rarely used to perform ground state AIMD, even though there is an ongoing  interest 
in unifying these two approaches \cite{Selloni1987, Rubio2008, Rubio2009}. 

In any case, if metallic system are treated, due to the fact that CP states are strictly not KS eigenstates, Eq.~(\ref{PastoreBuda}) is identical zero and either a thermostat for the electronic degrees of freedom to counterbalance the exchange of energy \cite{Bloechl1992, Bloechl2002}, or an extended functional using fractional occupation numbers \cite{Mermin1965, Gillan1989, Alavi1994, Marzari1997} is required.

In the end, drawing a proper conclusion if either BOMD or CPMD is to favor, turns out to be rather subtle \cite{Tangney2006} and depends largely on the definition of accuracy, as well as on the particular application.

\section{Second Generation Car-Parrinello Molecular Dynamics}


Even though DFT-based AIMD has been very successful in describing a large variety of physical phenomena, its rather high computational cost has limited the attainable length and time scales in spite of substantial progress \cite{APJ1992b, CarParrinelloUSPP, BloechlPAW1994, KresseFurthmueller1996b, Bernholc1996, Kaxiras1997, Lippert1999, Krack2000, LiuYarneTuckerman2003, Mostofi2003, Lilienfeld2004, Iannuzzi2006, Khaliullin2006, VandeVondeleHutter2012}. For a while it was believed that linear scaling methods \cite{Yang1991, GalliParrinello1992, Goedecker1999} could have offered a solution. Unfortunately, the crossover point at which linear scaling methods become advantageous has remained fairly large, especially if high accuracy is needed \cite{CeriottiKuehneParrinello2008, CeriottiKuehneParrinello2009, Richters2013}.
Thus, it would be very desirable to accelerate \textit{ab-initio} simulations with up to thousands of atoms, such that simulations as long as a few nanoseconds can be routinely performed, thus making completely new phenomena accessible to AIMD simulations. The direct BOMD approach, where the DFT functional is fully minimized at each MD time step, does not seem to offer much room for further improvement. For this reason recently another direction has been followed to improve the efficiency at current system sizes. In the spirit of CPMD \cite{CarParrinello1985}, some form of dynamics for the electronic degrees of freedom is implemented, which automatically keeps the system close to the instantaneous BO surface, but at variance to the original scheme in a localized orbital representation \cite{Tymczak1997, Iyengar2001, SharmaCar2003, ThomasTuckerman2004, HerbertHeadGordon2004, Schmid2006, Rapacioli2007}. The acceleration stems on the one hand from this more compact representation of the electronic wavefunctions, but is on the other hand mainly due to the ability to reduce or even fully bypass the aforementioned SCF cycle. Nevertheless, just like in CPMD, all of these techniques suffer from rather short integration time steps. 

\subsection{An Efficient and Accurate Car-Parrinello-like approach to Born-Oppenheimer MD}

The recently developed second-generation CPMD method of K\"uhne et al. combines the accuracy and long time steps of BOMD with the efficiency of CPMD \cite{KuehneParrinello2007}. In this Car-Parrinello-like approach to BOMD, the original fictitious Newtonian dynamics of the electronic degrees of freedom is substituted by an improved coupled electron-ion dynamics that keeps the electrons very close to the instantaneous BO surface and does not require the introduction of an artificial mass parameter. The superior efficiency of second-generation CPMD, which, depending on the systems, varies between one to two orders of magnitude, has been demonstrated for a wide range of different applications \cite{Caravati2007, KuehneKrackParrinello2009, Caravati2009a, Caravati2009b, Camellone2009, Cucinotta2009, KuehneJung2011, Luduena2011a, Caravati2011, Luduena2011b, Pascal2012, KuehneKhaliullin2013}. 

Within mean-field electronic structure theories, such as Hartree-Fock or KS-DFT, the electronic wavefunction is represented by a rectangular coefficient matrix $\mathbf{C}$ that has the dimension of the basis set size $M$ times the number of occupied states $N$ and includes the expansion coefficients of Eq.~(\ref{LCAO}). Because we will consider here the general case, where the orbitals are expanded in a non-orthogonal basis set, the $M\times M$ overlap matrix $\mathbf{S}$ is different from unity. The contra-covariant density matrix can then be written as $\mathbf{PS}$, where $\mathbf{P}=\mathbf{CC}^{T}$ is the one-particle density kernel. 
However, similar to the holonomic orthonormality condition of the orbitals, the $\mathbf{P}$ matrix must obey the equivalent idempotency condition $\mathbf{P}=\mathbf{PSP}$ that is due to the fermionic nature of electrons, which compels the wavefunction to be antisymmetric in order to meet the Pauli exclusion principle. The notion that the dynamics of $\mathbf{PS}$ is much smoother than the one of the more widely varying wavefunctions immediately suggests to propagate $\mathbf{PS}$, rather than $\mathbf{C}$ as in CPMD.


\subsection{Coupled Electron-Ion Dynamics}

At variance to CPMD, where the fictitious dynamics of the electrons is derived from the modified Lagrangian of Eq.~(\ref{CPlagrangian}), in the second-generation CPMD method the improved coupled electron-ion dynamics is directly specified in terms of an predictor-corrector integrator scheme \cite{KuehneParrinello2007}. Since, contrary to the evolution of the nuclei nuclei, accuracy is crucial for the short-term integration of the electronic degrees of freedom, a highly accurate yet efficient algorithm is essential. On this account we have selected the always stable predictor-corrector (ASPC) method of Kolafa \cite{Kolafa2004, Kolafa2005}, though other choices are equally possible \cite{MartynaTuckerman1995, Niklasson2007, Niklasson2009}. 

Adapting Kolafa's method to this particular case, we write the predicted wavefunction at time $t_{n}$ in terms of the $K$ previous $\mathbf{PS}$ matrices as
\begin{eqnarray}
\mathbf{C}^{p}\left( t_{n}\right) &\cong& \sum_{m=1}^{K}(-1)^{m+1}m%
\frac{{\binom{2K}{K-m}}}{{\binom{2K-2}{K-1}}}\underbrace{\mathbf{C}(t_{n-m})%
\mathbf{C}^{T}(t_{n-m})}_{\mathbf{P}(t_{n-m})} \nonumber \\
&\times& \mathbf{S}(t_{n-m}) \mathbf{C}\left( t_{n-1}\right).
\label{ASPC}
\end{eqnarray}%
that employs the extrapolated $\mathbf{PS}$ matrix as an approximate projector onto the occupied subspace $\mathbf{C} \left( t_{n-1}\right)$. In this way, we take advantage of the the fact that the physically relevant $\mathbf{PS}$ matrix evolves much more smoothly and is therefore much easier to predict than $\mathbf{C}$. The modified predictor is followed by a corrector step to minimize the error and to further reduce the deviation from the instantaneous ground state. Considering that this scheme was originally devised to deal with classical polarization, special attention must be paid that during the evolution the idempotency condition is always satisfied. Because of that, the modified corrector 
\begin{eqnarray}
\mathbf{C}\left( t_{n}\right)  &=&\omega \, \text{MIN}\left[\mathbf{C}^{p}
\left( t_{n}\right)\right] +(1-\omega )\mathbf{C}^{p}\left( t_{n}\right), \nonumber \\
\mbox{where~}\omega  &=&\frac{K}{2K-1}~\text{and}~K \ge 2
\end{eqnarray}%
involves the evaluation of only a single preconditioned electronic gradient $\text{MIN}\left[\mathbf{C}^{p}(t_{n})\right]$, using an idempotency conserving minimization technique, to calculate the electronic force. This predictor-corrector scheme leads to an electron dynamics that is accurate and time reversible up to $\mathcal{O}(\Delta t^{2K-2})$, while $\omega$ is chosen to guarantee a stable relaxation towards the instantaneous ground state.

Alternatively, the predictor can also be repeatedly applied, in which case the ground state is even more closely approached, though at the cost of additional electronic force calculations. However, the efficiency of this method is such that the electronic ground state is very closely approached within just one preconditioned electronic gradient calculation per AIMD step, so that in general this is not necessary. Therefore, the second-generation CPMD method not only entirely avoids the SCF cycle, but also the 
iterative wavefunction optimization, i.e. not even a single diagonalization step is required, while at the same time remaining very close to the BO surface and allowing for $\Delta t$ to be as large as in standard BOMD.

\subsection{Electronic Forces by Orbital Transformations}

Given that the energy is invariant under a unitary transformation within the subspace of occupied orbitals $\mathbf{C}$, it must be ensured that this gauge transformation is not strongly changed by $\text{MIN}\left[\mathbf{C}^{p}(t_{n})\right]$, as in this case continuity between the $\mathbf{C}$'s may be lost. Furthermore, the minimization scheme must be very efficient in bringing the electronic degrees of freedom as close as possible to the instantaneous ground state and at the same time preserves the idempotency condition of the density matrix. For these reasons, the orbital transformation (OT) method has been chosen \cite{OT2003}. Inspired by the form of the exponential transformation an auxiliary variable $\mathbf{X}$ is introduced, to parameterize the occupied orbitals
\begin{equation}
\mathbf{C} \left( \mathbf{X}\right) =\mathbf{C}^{p}\left( t_{n}\right) \cos
\left( \mathbf{U}\right) +\mathbf{X}\mathbf{U}^{-1}\sin \left( \mathbf{U}%
\right), 
\end{equation}%
where  $\mathbf{U=}\left( \mathbf{X}^{T}\mathbf{SX}\right) ^{1/2}$ and the variable $\mathbf{X}$ has to obey the linear constraint $\mathbf{X}^{T}\mathbf{SC}^{p}\left( t_{n}\right) =0$. Under this condition, $\mathbf{C} \left(\mathbf{X}\right)$ leads to an idempotent density matrix for any choice of $\mathbf{X}$, provided that the reference orbitals $\mathbf{C}^{p}\left( t_{n}\right)$ are orthonormal. Thus, any finite step along the preconditioned gradient direction will exactly fulfill the idempotency constraint by construction. Because of the linear constraint the minimization with respect to $\mathbf{X}$ is performed in an auxiliary tangent space. As this space is linear, no curved geodesics must be followed, as is the case for variables such as $\mathbf{C}$ that are nonlinearly constrained. In this way, large minimization steps can be taken, especially if a good preconditioner is used \cite{Gan2000}. In fact, using an efficient, idempotency conserving direct minimizer such as OT is decisive for the success of this approach. Since the ASPC integrator only approximately preserves the idempotency constraint, it occasionally has to be explicitly enforced by individual purification iterations \cite{PalserManolopoulos1998}.

\subsection{Total Energies and Nuclear Forces}

Having obtained the new wavefunction we are now able to evaluate the energy and the nuclear forces that are derived from the following approximate energy functional:
\begin{eqnarray}
E_{\text{PC}}[\rho^{p}] &=&{\text{Tr}}\left[ \mathbf{C}^{T}H[\rho
^{p}] \, \mathbf{C}\right] -\frac{1}{2}\int {d\mathbf{r}\int {d\mathbf{r^{\prime} }}%
\, \frac{\rho ^{p}(\mathbf{r})\rho ^{p}(\mathbf{r}^{\prime })}{|\mathbf{r}-%
\mathbf{r}^{\prime }|}}  \notag \\
&-&\int {d\mathbf{r} \,V_{\text{XC}}[\rho }^{p}{]}\rho ^{p}+E_{\text{XC}}[{\rho }%
^{p}]+E_{II} \mbox{,}
\label{DensityFunctional}
\end{eqnarray}%
where $\rho ^{p}(\mathbf{r})$ is the electron density associated with $\mathbf{C}^{p}(t_{n})$. $E_{\text{PC}}[\rho]$ can be thought of as an approximation to the Harris-Foulkes functional \cite{Harris1985, FoulkesHaydock1989} and maintains the predictor-corrector flavor inherent to the second-generation CPMD method. The validity of $E_{\text{PC}}[\rho]$ depends only on the efficiency of the minimizer and on the accuracy of the propagation scheme for the electronic degrees of freedom. The nuclear forces are calculated by evaluating the analytic gradient of $E_{\text{PC}}[\rho]$ with respect to the ionic coordinates. However, since $\Delta{\rho} = \rho - \rho^{p} \neq 0$, besides the usual Hellmann-Feynman and Pulay forces an extra term appears: 
\begin{equation}
-\int {d\mathbf{r}\left\{ \left[ \left( \frac{\partial V_{\text{XC}}[{\rho}^{p}]%
}{\partial {\rho}^{p}}\right) \Delta {\rho}+V_{\text{H}}[\Delta {\rho}]%
\right] \left(\nabla _{\mathbf{R}_I}{\rho }^{p} \right) \right\} } \mbox{,}
\label{FNV}
\end{equation}%
where $\rho$ is the corrected electron density, which corresponds to $\mathbf{C}(t_{n})$, while ${\rho}^{p}$ is the predicted electron density calculated from $\mathbf{C}^{p}(t_{n})$. Using variational density functional perturbation theory \cite{PutrinoSebastianiParrinello2000, BenoitSebastianiParrinello2001}, Eq.~(\ref{FNV}) can be efficiently computed very similar to employing the coupled-perturbed KS scheme. 
However, due to the fact that usually only a single preconditioned minimization step is performed, $\mathbf{C}(t_{n})$ is just an approximate eigenfunction of $H[\rho^{p}]$ within the subspace spanned by the finite basis set used. This leads to an insignificant error in the forces, provided that $\mathbf{C}(t_n)$ is very close to the ground state.

\subsection{Modified Langevin Equation}

Despite the close proximity of the electronic degrees of freedom to the instantaneous ground-state, the nuclear dynamics is dissipative, most likely because the employed electron propagation scheme is not symplectic. 
However, it is possible to rigorously correct for the dissipation by devising a modified Langevin equation, which in its general form reads as: 
\begin{equation}
M_{I}{\mathbf{\Ddot{R}}}_{I} = \mathbf{F}_{I}^{\text{BO}} - \gamma M_{I} {\mathbf{\Dot{R}}}_{I} + \mathbf{\Xi }_{I},
\label{LangevinEq}
\end{equation}
where $\mathbf{F}_{I}^{\text{BO}}$ are the correct but unknown BO forces, $\gamma$ a damping coefficient and $\mathbf{\Xi}_{I}$ an additive white noise that must obey the fluctuation-dissipation theorem $\left\langle \mathbf{\Xi}_{I}(0) \, \mathbf{\Xi}_{I}(t) \right\rangle = 2 \gamma M_{I} k_{B} T \delta(t)$ in order to correctly sample the canonical distribution. 

Presuming that the energy dissipation is indeed exponential, which had been shown to be an excellent assumption \cite{KuehneParrinello2007, Dai2009, Hutter2012}, it is possible to remedy this downward drift if we assume that the nuclear forces arising from our dynamics $\mathbf{F}_{\text{PC}} = - \nabla _{\mathbf{R}_I} E_{\text{PC}}[\rho^{p}]$ can be modeled as 
\begin{equation}
\mathbf{F}_{\text{PC}}=\mathbf{F}_{\text{BO}} - \gamma_{D} M_{I} {\mathbf{\Dot{R}}}_{I}, \label{TdKmodel}
\end{equation}
where $\gamma_{D}$ is an intrinsic, yet unknown damping coefficient to mimic the dissipation. By substituting Eq.~(\ref{TdKmodel}) into Eq.~(\ref{LangevinEq}) the following modified Langevin-like equation is obtained:
\begin{equation}
M_{I}{\mathbf{\Ddot{R}}}_{I} = \mathbf{F}_{I}^{\text{BO}} + \mathbf{\Xi }_{I},
\label{modLangevinEq}
\end{equation}

In other words, if one knew the unknown value of $\gamma_{D}$ it would nevertheless be possible to guarantee an exact canonical sampling of the Boltzmann distribution (in spite of the dissipation) by simply augmenting Eq.~(\ref{TdKmodel}) with $\mathbf{\Xi}_{I}$ according to the fluctuation-dissipation theorem. Fortunately, the inherent value of $\gamma_{D}$ does not need to be known \textit{a priori}, but can be bootstrapped by requiring that $\gamma_{D}$ is such as to generate the correct average temperature, as measured by the equipartition theorem. More precisely, we vary the unknown value of $\gamma_{D}$ in a preliminary run on-the-fly using a Berendsen-like algorithm until eventually Eq.~(\ref{EquipartitionTheorem}) is satisfied \cite{KrajewskiParrinello2006a}. Even though this can be somewhat laborious, but once $\gamma_{D}$ is fixed, very long and accurate AIMD simulations can be routinely performed at a much reduced computational cost.

\subsection{Illustrative Examples: Liquid Silicon, Silica and Water}

For the purpose of demonstrating this new approach, we have implemented it in the mixed Gaussian Plane Wave \cite{Lippert1997} code \textsc{Quickstep} \cite{Krack2004, VandeVondele2005}, which is part of the publicly available suite of programs CP2K \cite{CP2K}. In order to illustrate that this method works well irrespective of band-gap, system size and type, calculations on liquid metallic silicon, silica and water are presented. All of these systems are known to be very difficult, and are examples of liquid metals ($\text{Si)}$, complex and highly polarizable ionic liquids ($\text{SiO}_{\text{2}})$, as well as hydrogen bonded fluids ($\text{H}_{\text{2}}$O). The fact that the simulations have been performed in the liquid phase at 3000~K,  3500~K and 325~K respectively, leads to rapidly varying density matrix elements, thus making the propagation of the electronic degrees of freedom particularly challenging. Hence, the selected test cases can be considered as worst-case scenarios for any computational method.

All simulations have been performed at their experimental liquid densities using triple-zeta (TZV2P) basis sets, adequate density cutoffs, norm-conserving Goedecker-Teter-Hutter pseudopotentials \cite{GTH1996, Krack2005} and the generalized gradient approximation to the exact exchange and correlation functional \cite{PBE1996}. 
For simplicity the Brillouin zone is sampled at the $\Gamma $-point only, while Eq.~(\ref{modLangevinEq}) is integrated using the algorithm of Ricci and Ciccotti \cite{RicciCiccotti2003}, where the values for $\gamma_{D}$ turned out to be in the range of $10^{-4}~\text{fs}^{-1}$. The new $\mathbf{C}$'s are predicted using $K=4$ in Eq.~(\ref{ASPC}), which ensures time-reversibility up to $O(\Delta t^{6})$. 

\begin{figure}
\includegraphics[width=8.6cm]{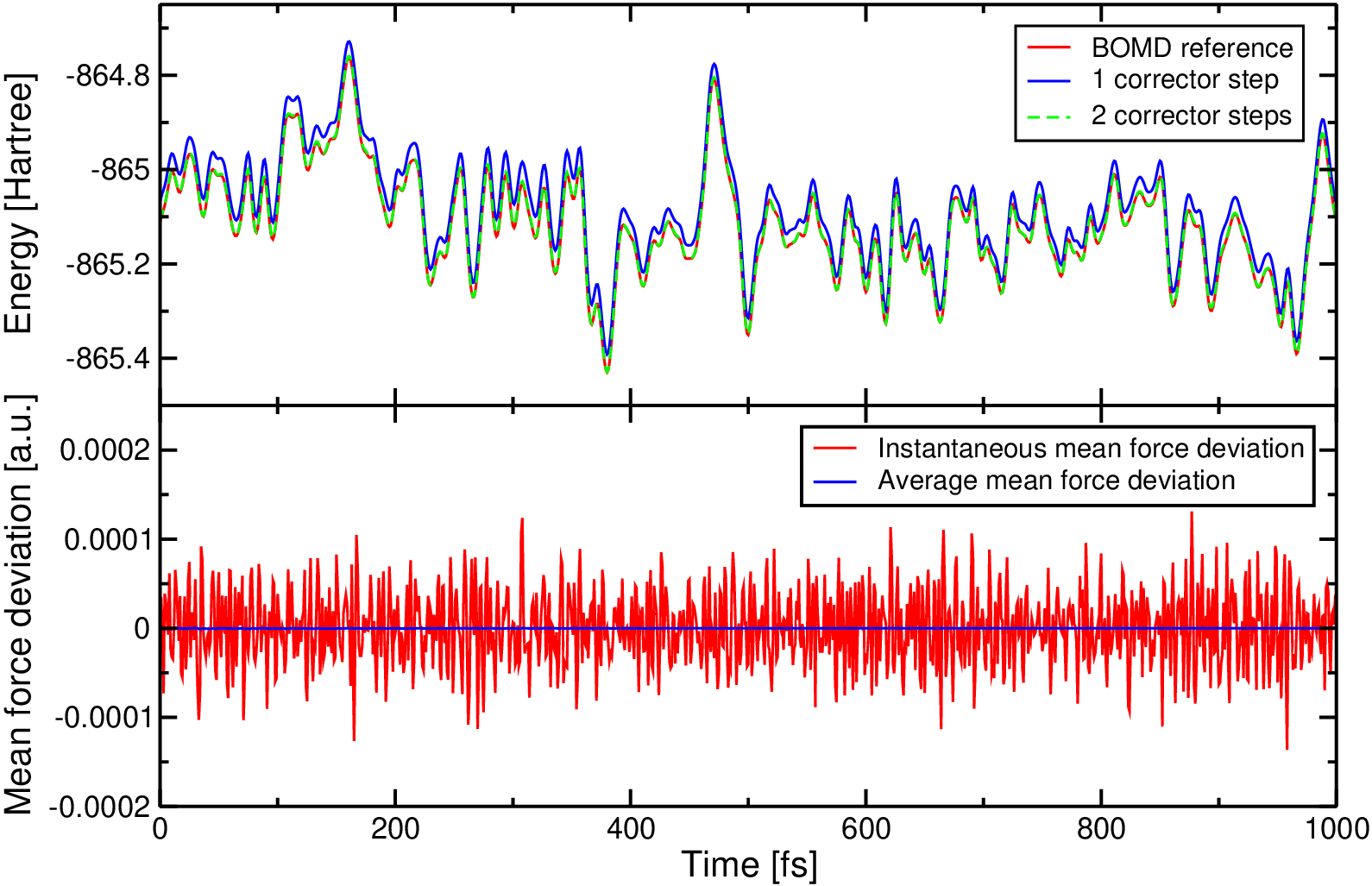}
\caption{Deviations from the BO surface of liquid $\text{SiO}_{\text{2}}$ in terms to total energies (upper panel) and mean force deviations (lower panel). The deviation in the energies corresponds to a constant shift of $4.16\times10^{-4}$ Hartree per atom for one corrector step and $3.5\times10^{-5}$ Hartree per atom for two corrector steps. The average mean force deviation is unbiased.} 
\label{RefTraj}
\end{figure}

First, the accuracy in terms of the energetic deviation from the BO surface is considered. As can be seen in FIG.~\ref{RefTraj} the energies are an upper bound to the electronic ground state and are displaced by a very small and approximately constant amount. It is also shown that, as already mentioned, the deviation from the BO surface can be even further reduced by increasing the number of corrector steps. In fact, it is actually possible to control the deviation from the BO surface by varying the number of corrector steps in order to achieve a preassigned accuracy level. However, in the following only simulations based on a single corrector step, i.e. only one preconditioned electronic gradient calculation, will be reported.

\begin{figure}
\includegraphics[width=8.6cm]{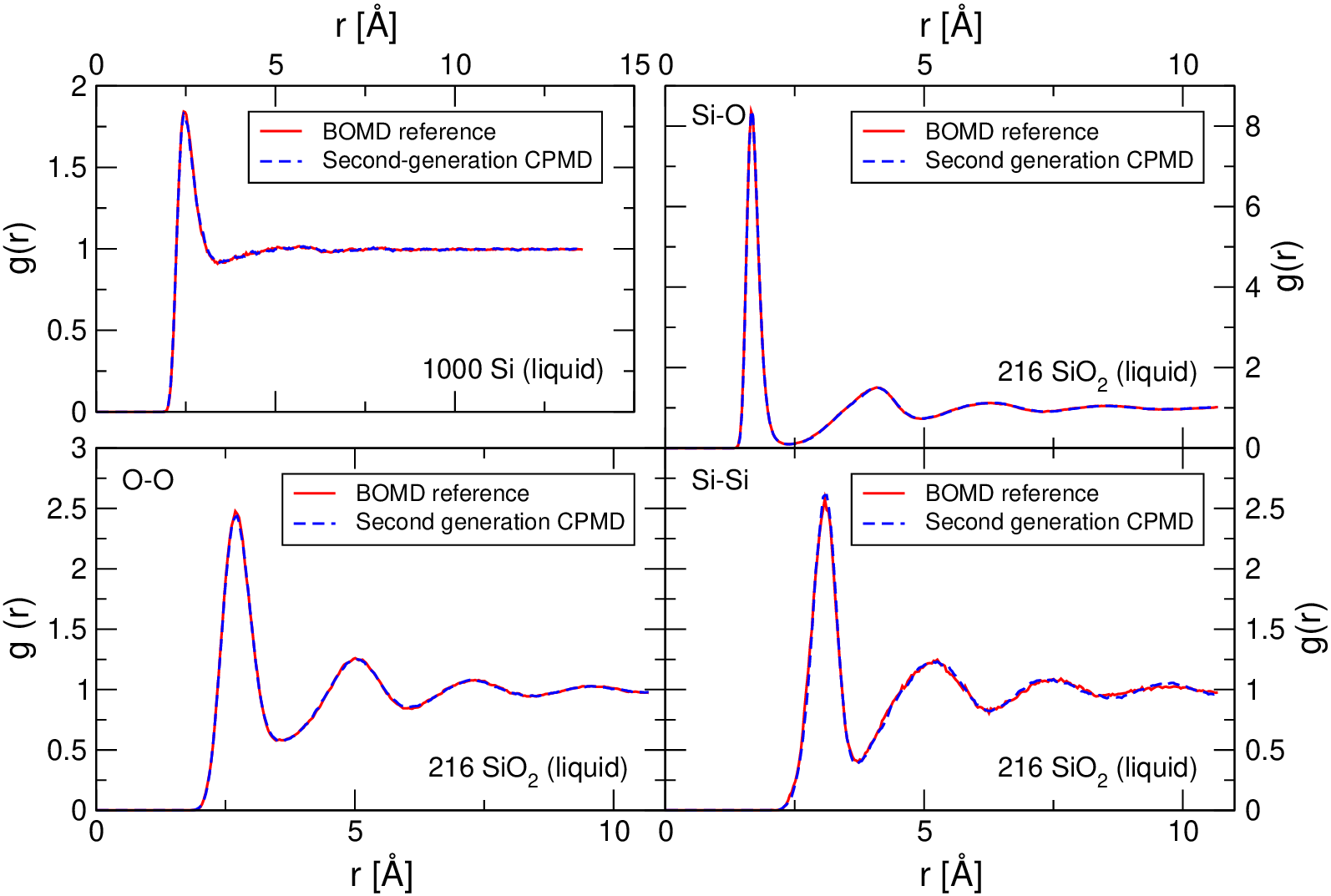}
\caption{Partial pair-correlation functions g(r) of liquid $\text{Si}$ (upper left panel) and liquid $\text{SiO}_{\text{2}}$ at 3000~K and 3500~K, respectively.}
\label{SiliconSilicaSquare}
\end{figure}

Nevertheless, let us now to turn to more realistic problems such as those shown in FIG.~\ref{SiliconSilicaSquare}.
Although these simulations have been performed with only a single corrector step, they are still amazingly close to the BOMD reference results. It should be emphasized that even in liquid Si, which is metallic and poses problems when using an ordinary CP scheme, a single corrector step is sufficient. This establishes the efficiency of this method, since only a single preconditioned gradient calculation with no additional minimization step has to be performed. The possible acceleration, in comparison with regular BOMD calculations, depends crucially on the system studied. In the undoubtedly difficult cases just presented a speed-up of two orders of magnitude compared to using a pure extrapolation scheme have been observed \cite{APJ1992b, VandeVondele2005}. For simpler problems still an increase in efficiency of at least one order of magnitude can be expected.

\begin{figure}
\includegraphics[width=8.6cm]{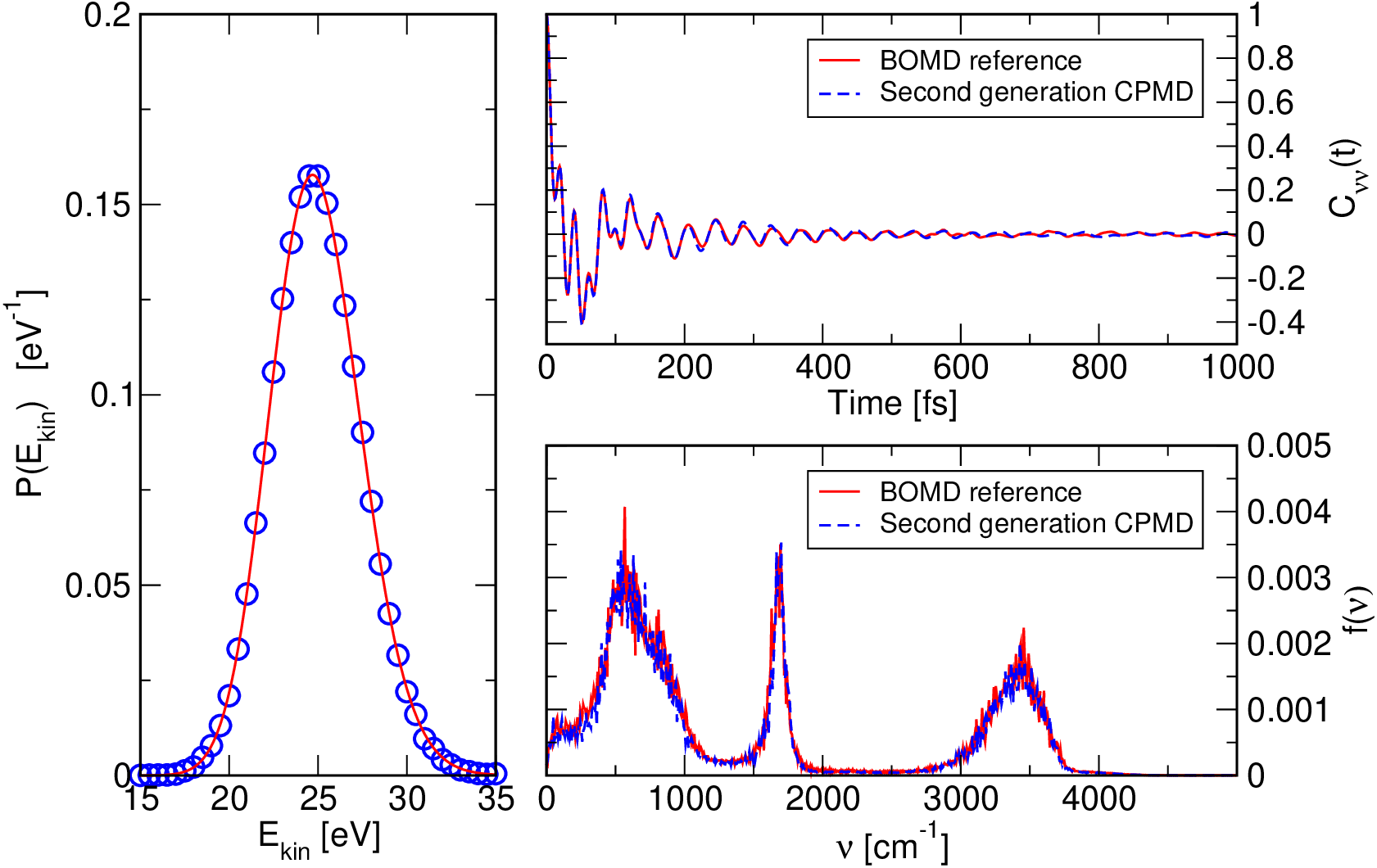}
\caption{The kinetic energy distribution calculated from a 1~ns trajectory of metallic liquid $\text{Si}_{\text{64}}$ (left panel). Velocity autocorrelation function (upper right) and its temporal Fourier transform (lower right) of 32~Water at 325~K. The unknown Langevin friction coefficient $\gamma_{D} \sim 10^{-8}~\text{fs}^{-1}$.}
\label{KinEk}
\end{figure}

In FIG.~\ref{KinEk} we present results, which demonstrate that also dynamical properties can be accurately calculated. To that extend the velocity autocorrelation function and its temporal Fourier transform at 325~K is shown. The results are in good agreement with accurate reference calculations and are consistent with experiment, as well as \textit{ab-initio} all-electron calculations \cite{Krack2000}, showing that in spite of the stochastic nature of Eq.~(\ref{modLangevinEq}) dynamical properties can also be simulated. This implies, that also chemical reactions and even non-equilibrium processes can be treated. In the same picture 
it is explicitly verified that the previous assumptions are justified, and indeed a canonical sampling is performed, by showing that the kinetic energy distribution is Maxwellian distributed. To this end, a 64 atom liquid Si simulation is carried out for as long as 1~ns, to reduce the noise and to ensure a proper sampling of the relevant kinetic energy distribution tails.

Due to space considerations only a fraction of the systems studied are reported here. Nevertheless, in all cases second-generation CPMD has proven to be very accurate and the gain in speed has always been remarkable \cite{KuehneKrackParrinello2009, Cucinotta2009, KuehneJung2011, Luduena2011a, Luduena2011b, Pascal2012, KuehneKhaliullin2013}. Furthermore, structure relaxation via dynamic annealing and geometry optimization have also been successfully performed \cite{Caravati2007, Caravati2009a, Camellone2009, Caravati2011}. Thanks to this development it is now possible to perform AIMD simulations on medium-sized systems up to a few thousands of atoms for as long as a couple of nanoseconds, thus making a new class of problems accessible to \textit{ab-initio} simulations.  



\section{Acknowledgments}

The author would like to thank Matthias Krack and Michele Parrinello for various fruitful discussions, as well as the whole CP2K-Team. In addition he is indebt to the Graduate School of Excellence MAINZ and the IDEE project of the Carl-Zeiss Foundation for financial support. The generous allocation of computer time and support from CSCS Manno, the ICT Services of ETH Z\"urich and the DEISA consortium is kindly acknowledged. 


\end{document}